# Improving the Electro-Optical Properties of MoS$_2$/rGO Hybrid Nanocomposites Using Liquid Crystals

A. Vasil'ev[1], Y. Melikyan[1], M. Zhezhu[1], V. Hayrapetyan[1], M. S. Torosyan[1], D. A. Ghazaryan[3, 4], M. Yeranosyan[1, 2], H. Gharagulyan[1, 2] *

[1] *Innovation Center for Nanoscience and Technologies, A.B. Nalbandyan Institute of Chemical Physics NAS RA, Yerevan 0014, Armenia*

[2] *Institute of Physics, Yerevan State University, Yerevan 0025, Armenia*

[3] *Moscow Center for Advanced Studies, Moscow 123592, Russia*

[4] *Laboratory of Advanced Functional Materials, Yerevan State University, Yerevan 0025, Armenia*

*Author to whom correspondence should be addressed: herminegharagulyan@ysu.am

**Abstract**

Hybrid systems of two-dimensional (2D) materials (such as graphene-family materials and 2D transition metal dichalcogenides) are attracting much attention due to their distinctive optoelectronic, thermal, mechanical, and chemical properties. The application perspectives of these materials in various fields further expand when enriching those with liquid crystals (LCs) primarily due to their enhanced tunability and functionality. In this study, we report on the hydrothermal synthesis of hybrid nanocomposites composed of MoS$_2$ and rGO and discuss tuning possibilities of their electro-optical properties by incorporating thermotropic LCs. In particular, we demonstrate that the incorporation of 5CB LC increases the sensitivity and charge storage efficiency of the hybrid nanocomposites. In addition, we also present the responsivity, detectivity, and response time properties of the hybrid nanocomposites of MoS2/rGO, both with and without the inclusion of nematic LCs. Furthermore, we demonstrate that the system exhibits a 5CB-induced photocurrent switching effect. We believe the findings will open new doors for applications of these materials in optoelectronics and photonics.

**Abbreviations:** rGO: reduced graphene oxide; TMDC: transition metal dichalcogenide; MoS$_2$: molybdenum disulfide; LC: liquid crystal; S: sensitivity; R: responsivity; D: detectivity; PL: photoluminescence; 5CB: 4-cyano-4′-pentylbiphenyl.

## Introduction

It is difficult to imagine modern technologies without two-dimensional (2D) materials, including metallic, semimetallic, superconducting, semiconducting, and insulating materials, such as black phosphorus and hexagonal boron nitride [1]. Recently, there has been a significant focus on graphene-family materials (such as graphene, graphene oxide (GO), and reduced GO (rGO)) and 2D transition metal dichalcogenides (TMDCs, e.g., molybdenum disulfide (MoS$_2$)) due to their outstanding optoelectronic, thermal, mechanical, and chemical properties [2]. However, integrating these materials into modern optoelectronic and photonic basic circuits is restricted by scaling, productivity, and tunability challenges. The tuning of the optical properties of 2D materials by external stimuli, including electric field, mechanical strain, magnetic field, acoustic wave, and thermal heating, partially solves those limitations and expands the scale of their applications [3].

Lyotropic liquid crystalline (LC) phases of 2D materials also give additional tuning possibilities since, in LCs, electrical conductivity and optical and thermal properties may vary with the orientation of LC directors [4, 5]. Conversely, using 2D materials in thermotropic LCs can also improve their electro-optical properties [6]. The ability of LCs to spontaneously arrange external particles into regular geometric design shows a diversity of self-assembled structures, such as linear chains, anisotropic clusters, 2D hexagonal lattices at interfaces, arrays of defects, particle-stabilized gels, and cellular soft-solid structures [7]. Among thermotropic LCs, the 4-cyano-4′-pentylbiphenyl (5CB) offers an advantage of great electro-optical responsiveness [6]. In particular, ground-state electronic properties of 5CB on 2D materials, such as graphene, hexagonal boron nitride, and phosphorene, are studied in [8], and tuning of the PL spectrum of $MoS_2$ with 5CB phase is discussed in [9].

$MoS_2$/rGO hybrid structures are of interest since they offer a large surface area and improve electronic conduction efficiency [10]. There are many methods for the synthesis of these structures, such as hydrothermal process [11, 12], microwave annealing-assisted [13] and PVP-assisted ultrasonication methods [14], and microwave-assisted solvothermal synthesis [15]. Those can be used in diverse applications, such as for energy generation and storage applications [16], supercapacitors [17, 18], alcohol fuel cells [19], dye-sensitized solar cells [20, 21], wave absorption and stealth camouflage techniques [22, 23], electromagnetic interference shielding [24], field effect transistors [25], and photodetectors [26, 27].

Graphene-family materials and $MoS_2$ exhibit distinctive optoelectronic features appropriate for manufacturing 2D material-based optoelectric devices [28]. Unlike GO, $MoS_2$ has a thickness-dependent bandgap, tunable work function, high absorption coefficient, and strong light-matter interaction [10]. $MoS_2$/rGO hybrid structures show improved photodetection performance compared to constituent components [29]. The enhanced performance might be credited to the creation of numerous $MoS_2$/rGO interfaces, which augment light absorption efficiency and facilitate rapid charge transfer, leading to significantly elevated photocurrent and photosensitivity. These structures are intensively used in photodetectors and flexible electronics [30] due to their unique properties, such as the ability to operate in the full range of visible light, high polarization sensitivity, fast photoresponse, and high spatially resolved imaging.

In this study, we report on tuning possibilities of electro-optical properties of hydrothermally synthesized $MoS_2$/rGO hybrid nanocomposite by 5CB LC. Specifically, we focus on enhancing the sensitivity and charge storage properties of the hybrid nanocomposites. For this, voltammetry measurements under chopped light conditions are conducted using a white light source spanning wavelengths from 420 nm to 730 nm. Furthermore, the photocurrent switching effect is observed when incorporating 5CB into the hybrid nanocomposites, which preserves a large range of light intensities from 15 mW/cm$^2$ to 65 mW/cm$^2$. Besides, responsivity, detectivity, photocurrent rise, and decay times for the hybrid nanocomposites, both with and without the presence of 5CB, are also estimated.

## 1. Materials and Methods

**Materials.** All the chemicals used in the experiments were purchased from Sigma-Aldrich Chemical Co.

### 1.1. Synthesis of MoS$_2$/rGO hybrid nanocomposites

The 40 mL solution with $(NH_4)_6Mo_7O_{24} \cdot 4H_2O$ as Mo source and thiourea CS $(NH_2)_2$ as S source with Mo:S mole ratio of 1:4 was mixed with a 40 ml solution of electrochemically synthesized GO [31]. Then, the mixture was transferred into a PPL-lined autoclave. The reactor vessel was heated to 220°C and maintained at 8 hours. After cooling to room temperature, the black suspension was filtered and washed with water, ethanol, and acetone before being air-dried. The rGO was added as 1% mass to MoS$_2$.

### 1.2. Characterization

Crystallographic data for the MoS$_2$/rGO hybrid nanocomposites was acquired through XRD analysis using a MiniFlex instrument from Rigaku. Chemical compound analysis of the mentioned material was performed using FTIR-ATR spectrometry with a Spectrum Two device from PerkinElmer. The bond types and hybridization of the MoS$_2$/rGO were identified using Raman spectroscopy, employing an XploRA PLUS instrument from HORIBA. The chemical composition of both GO and the MoS$_2$/rGO hybrid nanocomposites was determined via HR XPS analysis conducted on a KRATOS Axis Supra+ instrument from Shimadzu. Optical characterization of these materials was done using PL spectroscopy with a Cary Eclipse Fluorescence Spectrometer from Agilent. Additionally, the optical properties of the MoS$_2$/rGO were investigated using UV-Vis spectrophotometry, utilizing a Cary 60 instrument from Agilent. The morphology and chemical composition of the aforementioned structure were analyzed using scanning electron microscopy (SEM) Prisma E from Thermo Fisher Scientific, as well as an image-corrected transmission electron microscopy (TEM) Titan 60-300 Themis, operated at 300 kV and equipped with a high-annular angular dark-field detector for scanning TEM imaging (STEM-HAADF) and a Super-X detector for energy-dispersive X-ray spectroscopy elemental mapping in STEM mode (STEM-EDX). The measurements of electro-optical properties of MoS$_2$/rGO and MoS2/rGO-5CB nanocomposites were conducted by Zahner photoelectrochemical workstation.

## 2. Results and Discussion

### 2.1. Structural Analysis of MoS$_2$/rGO hybrid nanocomposites

A comprehensive spectral and morphological study of the synthesized material was performed. In particular, the TEM image of MoS$_2$/rGO hybrid nanocomposite's morphology is shown in Fig. 1(a). It reveals that the MoS$_2$ nanoflowers have grown almost uniformly and anchored on wrinkled rGO's basal plane. The hierarchical MoS$_2$/rGO sandwich composite was formed eventually. A fast Fourier transform (FFT) and the selected-area electron diffraction (SAED) analysis were done to characterize the structure in detail. FFT measurements indicate the presence of 0.28 nm *d*-spacing values in the MoS$_2$ nanosheets, which corresponds to the (100) lattice spacing of the MoS$_2$ hexagonal phase (see Fig. 1(b)) [32]. However, there are parts in the structure where the interlayer spacing is about 0.41 nm. This is attributed to the intercalation, osmotic expansion, defects, and chemical functionalization of MoS$_2$ [33, 34]. The thickness of MoS$_2$ nanosheets was also estimated from the TEM image, which was approximately 0.65 nm per layer (when averaged over 10 layers). The TEM image also clearly shows rGO's atomic structure of 1 nm hexagon side length (zoomed from the selected area). Fig.1(c) illustrates the SAED distinct ring pattern of the MoS$_2$/rGO hybrid structure, where circles correspond to the rGO (the radius of the first circle is equal to half of the hexagon) and spots to the MoS$_2$, confirming the formation of highly crystalline nanostructure of MoS$_2$, which is similar to the crystalline structures of pure MoS$_2$ and consistent with the XRD results [35]. The EDX spectra obtained from the sample showed characteristic energy lines of C, Mo, S, and O elements (see Fig. 1(d)). In contrast, the elemental maps showed the distribution of the above-mentioned elements across the material surface (see Fig. S1 (b) in the Supplementary Materials). The atomic percentages for the hybrid nanocomposite were as follows: C, 22 %; Mo, 26 %; S, 52 % (oxygen contribution is not considered). At first, a fast screening was done to obtain a wide spectrum scan of the XPS spectrum of MoS$_2$/rGO hybrid nanocomposite (see Fig. 1 (e)). The existence of C$_{1s}$, O$_{1s}$, S$_{2p}$, and Mo$_{3d}$ implies the successful preparation of a MoS$_2$/rGO hybrid nanocomposite system. It is worth mentioning that in the XPS wide spectrum, the Si peak corresponds to the substrate. Here, we obtain that the O/C ratio calculated according to [36] for the MoS$_2$/rGO hybrid nanocomposite is 0.216, while it was 0.235 for pristine GO, indicating a successful reduction of GO. Figs. 1(f-i) present fitted and original spectra for MoS$_2$/rGO nanocomposite's elements, namely, Mo$_{3d}$, S$_{2p}$, C$_{1s}$, and O$_{1s}$, respectively. The high-resolution spectrum (see Fig. 1 (f)) of Mo$_{3d}$ exhibits two distinct peaks at 229.4 eV (Mo$_{3d\ 5/2}$) and 232.6 eV (Mo$_{3d\ 3/2}$), indicating the presence of the Mo (IV) oxidation state within MoS$_2$. An additional minor peak at 226.6 eV is attributed to S$_{2s}$, showing the presence of the S-Mo-S bond in MoS$_2$. Further, a comparatively minor peak located at 235.32 eV is assigned to Mo$_{3d}$, corroborating the presence of the Mo (VI) oxidation state. In the S$_{2p}$ spectrum (see Fig. 1(g)), two prominent peaks at 163.46 eV and 162.26 eV are identified as binding energies corresponding to S$_{2p\ 1/2}$ and S$_{2p\ 3/2}$, respectively, characteristic to 2*s* orbital. The low intensity of the peak at 235.32 eV indicates negligible amount of MoO$_3$ in the hybrid nanocomposites.

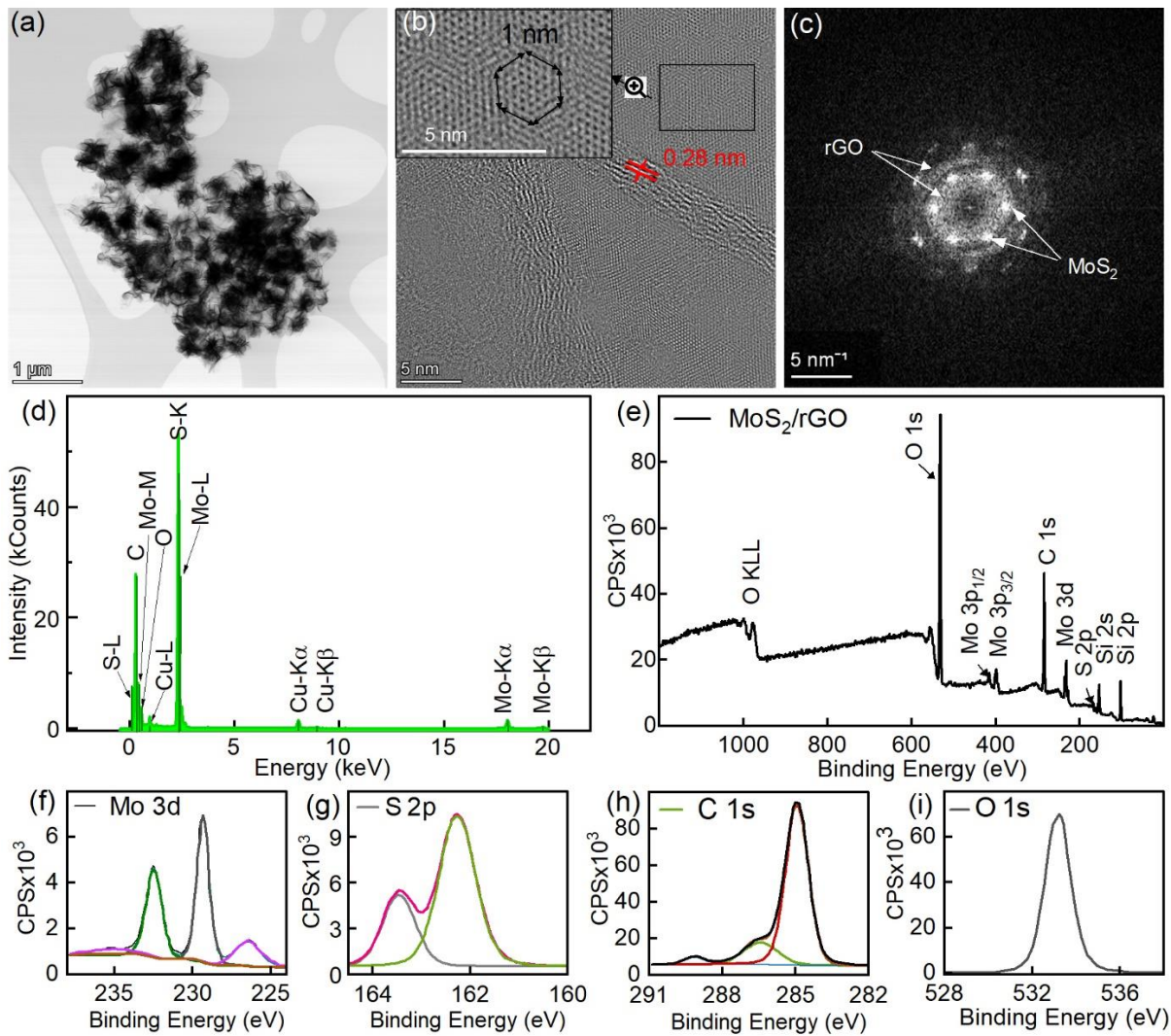

Fig. 1. Characterization of the synthesized MoS$_2$/rGO hybrid nanocomposites. (a) TEM image of MoS$_2$/rGO morphology. (b) FFT measurements of the TEM image. The inset shows the interlayer spacing of MoS$_2$ nanosheets and the hexagon side length of the rGO atomic structure. (c) SAED diffraction pattern. (d) Analysis of chemical elements in the material collected from the EDX spectrum. (e) XPS spectra of MoS$_2$/rGO hybrid nanocomposite. (f-i) XPS survey scan spectra for MoS$_2$/rGO: (f) Mo$_{3d}$; (g) S$_{2p}$; (h) C$_{1s}$; (i) O$_{1s}$ peaks. Binding energies were referenced to the C$_{1s}$ main peak of carbon tape at 285 eV.

Table 1 represents XPS binding energy peak positions for the MoS$_2$/rGO hybrid nanocomposites and their comparison to GO.

Table 1. XPS binding energy peak positions for MoS$_2$/rGO and GO.

| MoS$_2$/rGO | | | | | GO | | |
|---|---|---|---|---|---|---|---|
| *Peak position (eV)* | | | | | *Peak position (eV)* | | |
| **C$_{1s}$** | sp$^3$C | 285 | | | **C$_{1s}$** | sp$^3$C | 285.02 |
| | C=O | 286.43 | **S$_{2p\ 3/2}$** | 162.26 | | C-O | 286.85 |
| | O-C=O | 289.13 | **S$_{2p\ 1/2}$** | 163.46 | | O-C=O | 289.16 |
| **O$_{1s}$** | | 533 | | | **O$_{1s}$** | | 532.91 |
| **S$_{2s}$** | | 226.6 | | | | | |

| | | | | | | |
|---|---|---|---|---|---|---|
| Mo (IV)$_{3d\ 5/2}$ | | 229.4 | Mo$_{3p\ 3/2}$ | 396 | | |
| Mo (IV)$_{3d\ 3/2}$ | | 232.6 | Mo$_{3p\ 1/2}$ | 414 | | |
| Mo (VI) | | 235.32 | | | | |

The optical properties of the MoS$_2$/rGO were determined using UV-Vis spectroscopy and the obtained spectrum is presented in Fig. 2(a). There are four peaks on the UV-Vis spectrum of MoS$_2$/rGO at 270 nm, 425 nm, 613 nm, and 670 nm. At 270 nm, rGO exhibits π–π* transition in the aromatic C–C bonds of the reduced graphene oxide [37]. The MoS$_2$ absorption spectrum shows three characteristic peaks. The broad absorption peak within the region of 400–450 nm consisting of *C* and *D* doublet emerges from the excitonic transitions. In comparison, the peaks at 613 nm and 670 nm correspond to the *B*-exciton and *A*-exciton peaks of the MoS$_2$, arising from the *P*, *Q*, and *K* points of the Brillouin zone, respectively [37]. The bandgaps of the MoS$_2$/rGO, both direct and indirect ones, were estimated from the UV-Vis spectrum based on the Tauc equation:

$$(\alpha h\nu)^{\frac{1}{n}} = C(h\nu - E_g), \tag{1}$$

where $\alpha$ and $h\nu$ are the absorption coefficient and photon energy, $C$ is a proportionality constant determined by the refraction index, $E_g$ is the bandgap value, and $n$ is the exponent [38]. For direct bandgap, $n = 1/2$ and $E_g$ is determined as the intercept of the tangent of the Tauc plot on the *x*-axis. As can be seen from Fig. 2, for the MoS$_2$/rGO, the direct bandgap is 1.7 eV (see Fig. 2 (b)) and the indirect bandgap is 2.2 eV (see Fig. 2(c)), respectively [39]. Fig. 2 (d) shows the PL spectra of the MoS$_2$ and MoS$_2$/rGO hybrid nanocomposite. The PL peaks were observed at 485 nm, 530 nm, 544 nm, 642 nm (exciton B), and 689 nm (exciton A) in the case of the excitation wavelength of 400 nm [39, 40]. Generally, these peaks give insights into the electronic bandstructure, excitonic properties, and defect states of the hybrid nanocomposite, which are essential for understanding its optical and optoelectronic behavior, namely, photogenerated electron-hole pairs, charge separation, and recombination processes [41]. As can be seen from PL spectra, the MoS$_2$/rGO demonstrates an enhanced capacity to absorb more photons and exhibit reduced radiation compared to pure MoS$_2$ (synthesized by the same hydrothermal method as MoS$_2$/rGO), which offers better conditions for photocurrent generation [10, 42]. Fig. 2(e) presents the FTIR-ATR spectrum of the MoS$_2$/rGO. The corresponding peak positions of the functional groups are tabulated below. Here, the peak at 579 cm$^{-1}$ and 868 cm$^{-1}$ are associated with vibration modes of Mo-S [24, 43]. The peaks at 1039 cm$^{-1}$ and 1135 cm$^{-1}$ are the characteristic bonds of the C-O group [44]. The FTIR-ATR spectrum shows the characteristic peaks at 1357 cm$^{-1}$, 1540 cm$^{-1}$, 1735 cm$^{-1}$, and 3425 cm$^{-1}$ related to the vibration modes of C-O-C, C=C, C=O and O-H of COOH group, respectively [10]. Stretchings at 2677 cm$^{-1}$ and 2905 cm$^{-1}$ correspond to C-H group [45]. The Raman spectrum of the MoS$_2$/rGO is shown in Fig. 2(f). For MoS$_2$, it exhibits two active vibrational modes $E_{2g}$ (double-degenerate vibration at 378 cm$^{-1}$) and $A_{1g}$ (nondegenerate at 403 cm$^{-1}$), corresponding to the in-plane and out-of-plane strains [46]. An energy value difference between them is 25 cm$^{-1}$, which reflects the multilayer structure of MoS$_2$. *D* and *G* breathing vibrations at 1344 cm$^{-1}$ and 1578 cm$^{-1}$, respectively, correspond to the defects and $sp_2$ hybrid carbon atoms in the MoS$_2$/rGO structure [47]. The intensity of *D* peak is almost the same as for *G* peak ($I_D/I_G = 1.017$), indicating the presence of a particular degree of defects in the samples. As evidenced from Fig. 2 (g), the XRD patterns of MoS$_2$/rGO nanocomposite exhibit characteristic peaks at 14.06°, 33.44°, 39.76°, 49.25°, 59.15°, and 69.53°, which correspond to predominate (002), (100), (102), (105), (110), and (210) planes of MoS$_2$, respectively, and for rGO at 22° ((002) plane) [48]. Some characteristic XRD peaks of the $MoS_2$/rGO are less intense and wider than those of pristine $MoS_2$ due to the overlapping of the rGO layers between $MoS_2$.

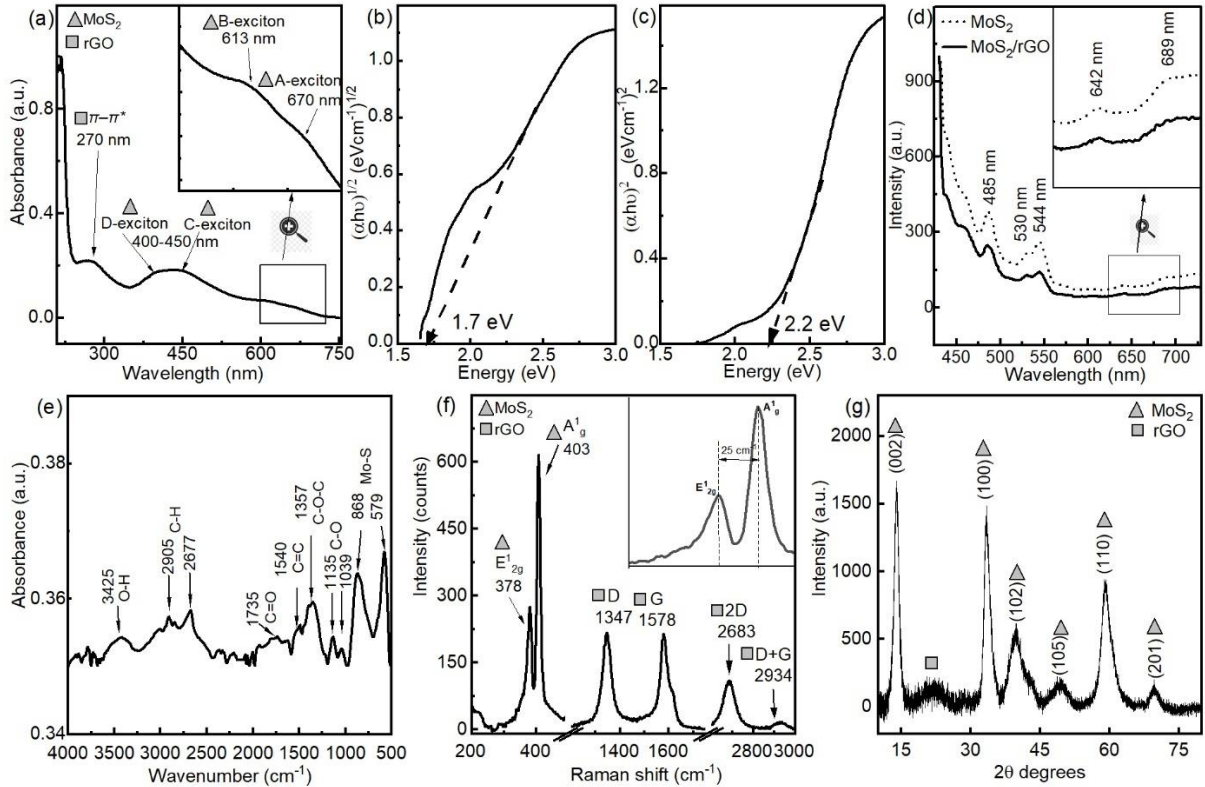

Fig. 2. Characterization of the synthesized $MoS_2$/rGO hybrid nanocomposites. (a) UV-VIS absorption spectrum of $MoS_2$/rGO. The inset shows three characteristic peaks for $MoS_2$ and 1 for rGO. Tauc plot of the optical absorption spectrum of $MoS_2$/rGO for direct (b) and indirect (c) bandgaps. (d) PL emission spectra (excitation wavelength: 400 nm; scan range: 430–750 nm; scan rate: 2 nm/s) for $MoS_2$ and $MoS_2$/rGO. (e) FTIR-ATR spectrum of $MoS_2$/rGO's functional groups. (f) Raman spectrum of $MoS_2$/rGO. The inset shows the $E_{2g}$ and $A_{1g}$ vibrational modes for $MoS_2$ and $D$ and $G$ peaks of rGO. (g) XRD spectrum of $MoS_2$/rGO. All measurements were conducted at an elevated temperature of 25° C.

## 2.2. Electro-optical properties of $MoS_2$/rGO and $MoS_2$/rGO-5CB hybrid nanocomposites

After synthesizing and characterizing the hybrid nanocomposites, we proceeded to tune their electro-optical properties with LCs. Chopped light voltammetry measurements were performed with white light source of 420 nm to 730 nm wavelengths and 18 mm diameter, which was controlled by a diaphragm. For measurements, interdigital electrodes (see Fig. S2 in the Supplementary Materials) with nickel contacts were used, where the corresponding mixtures of $MoS_2$/rGO and $MoS_2$/rGO-5CB were dropcasted on it with fixed mass ratios (for the case of the latter, 5CB was 5 % in the $MoS_2$/rGO mixture). It is worth mentioning that an active light feedback loop ensures a controlled and constant light intensity during the measurements in the system. As for all photoactive materials, four parameters, namely, photocurrent, sensitivity, responsivity, and detectivity, are discussed for comparison of electro-optical properties of $MoS_2$/rGO and $MoS_2$/rGO-5CB nanocomposites. Fig. 3 (a) represents the current-voltage characteristics for the structures at room temperature. As it can be seen from Fig. 3 (a), in the case of 5CB incorporation into the hybrid nanocomposites, the current-voltage dependency of $MoS_2$/rGO demonstrates nonlinearity due to the 5CB's nonlinear behavior of dielectric permittivity [49]. The electrical conductivity of thermotropic LCs is normally caused by ions [50], while it is known that MoS$_2$/rGO hybrid nanocomposites are *n*-type semiconductors [51]. Fig. 3(b) represents obtained typical photocurrent curves vs. time without dark current subtraction for MoS$_2$/rGO and MoS$_2$/rGO-5CB under 45 mW/cm$^2$ light illumination and 1 V electric field. From the phase intervals depicted in Fig. 3 (b), one can see that when the light is ON, the charge carriers are generated, while when it is OFF, the charge carriers decrease through direct or indirect recombination or trapping [52]. However, it is interesting to note that an LC-induced photocurrent switching was observed for the MoS2/rGO-5CB sample. Generally, a decrease in photocurrent under light illumination can occur due to saturation effects, recombination rates, transport limitations, and material properties, such as bandgap, surface conditions, purity, and electron mobility [53]. For our system, the emerging effect can be explained by the trapping centers existing in 5CB, although it can also happen due to the charge carriers' transport in the opposite direction. The photocurrent and dark current of MoS$_2$/rGO and MoS$_2$/rGO-5CB hybrid nanocomposites were measured in the absence of light and in the case of illumination with different intensities (see Fig. 3(c)). As observed from the measurement results, the photocurrent switching effect preserves for MoS$_2$/rGO-5CB in case of light intensities from 15 mW/cm$^2$ to 65 mW/cm$^2$, and the peaks for both samples are quite stable.

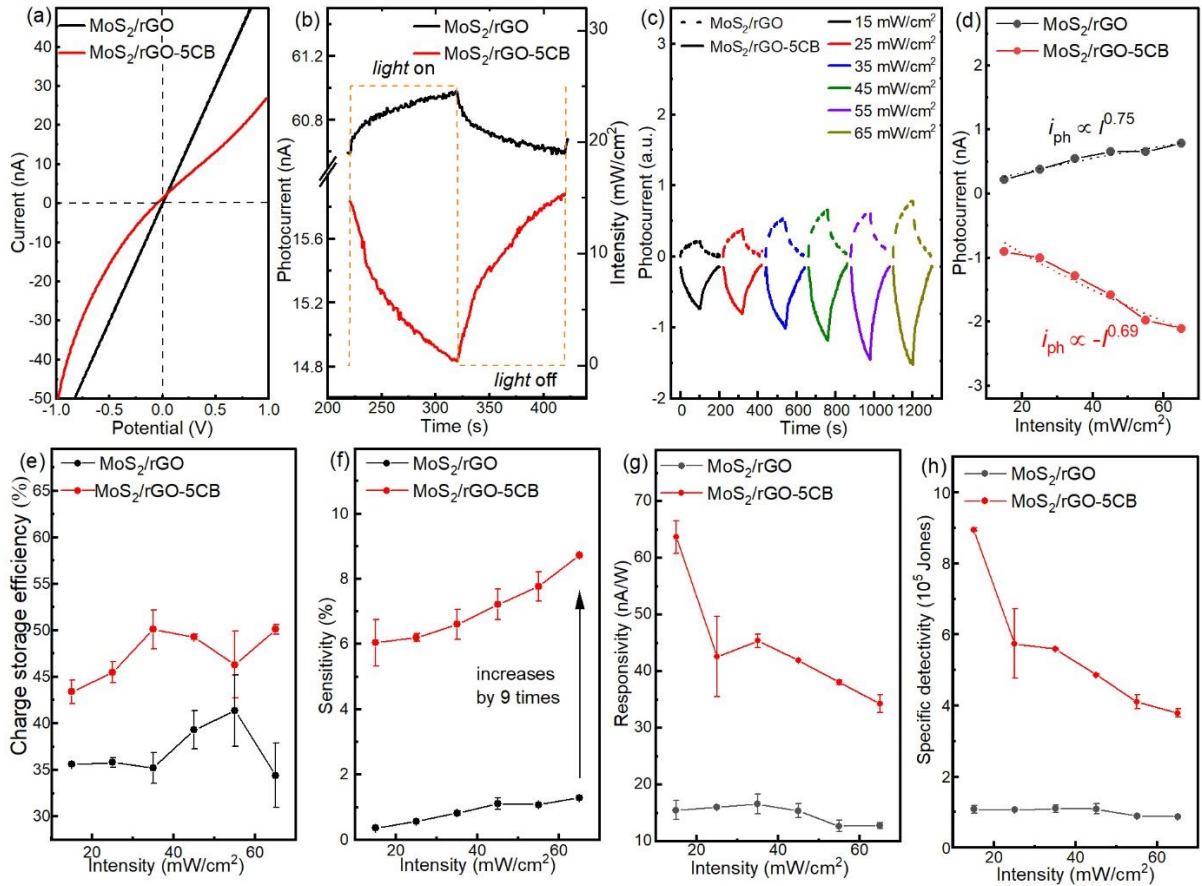

Fig. 3. Electro-optical characterization of MoS$_2$/rGO and MoS$_2$/rGO-5CB hybrid nanocomposites. (a) Current-voltage dependence in case of dark conditions. (b) Photocurrent curves vs. time in case of applied 45 mW/cm$^2$ light and 1 V electric fields. (c) Photocurrent vs. time for different intensities. (d) Fitted photocurrent vs. intensity according to the power law (2). (e) Charge storage efficiency vs. intensity. (f) Sensitivity vs. intensity. Responsivity (g) and specific detectivity (h) vs. intensity. The dependencies in (d–h) are fitted according to the power function: $y = ax^b$.

A quantitative estimation for the photocurrent rise and decay times of **the** MoS$_2$/rGO and MoS$_2$/rGO-5CB samples was done in the case of different light intensities (see Table 2) [54].

Table 2. Photocurrent rise and decay times for MoS$_2$/rGO and MoS$_2$/rGO-5CB.

| MoS$_2$/rGO | | | MoS$_2$/rGO-5CB | | |
|---|---|---|---|---|---|
| Intensity, mW/cm$^2$ | Rise time, s | Decay time, s | Intensity, mW/cm$^2$ | Rise time, s | Decay time, s |
| 15 | 64.10 | 65.37 | 15 | 75.44 | 70.77 |
| 25 | 64.30 | 70.44 | 25 | 73.18 | 72.54 |
| 35 | 58.41 | 66.52 | 35 | 79.53 | 76.20 |
| 45 | 68.38 | 66.65 | 45 | 75.63 | 74.87 |
| 55 | 72.25 | 64.35 | 55 | 71.67 | 70.45 |
| 65 | 71.65 | 70.00 | 65 | 67.63 | 76.57 |

The response time comparison for the samples shows that the capturing of photogenerated carriers increases the photocurrent decay time for the MoS$_2$/rGO-5CB sample, which confirms the existence of traps in the hybrid nanocomposite.

The photocurrent ($i_{ph}$) and the illumination power ($P$) are expressed by the following equation:

$$i_{ph} = a \cdot P^b, \qquad (2)$$

where $b$ is the dimensionless exponent of the power law and $a$ is a parameter related to the photodetector responsivity [54]. Conversely, the value of the exponent provides information on the presenting traps in the system. When exponent $b$ equals 1, the photocurrent depends linearly on the illumination power, indicating that the photodetectors are without traps. Fig. 3(d) demonstrates clearly that the exponent for the samples is less than 1, which means that the electrons will not only recombine with holes but will also be captured by traps. When the excitation light is turned OFF, the depletion of traps will delay the decrease in the concentration of charge carriers. As a result, this will lead to a slowdown in the process of increasing and decreasing photoconductivity [55, 56]. Fig. 3(e) depicts the dependence of the charge storage efficiency on intensity determined by the following expression:

$$\eta = Q_{us}/Q_{uo}, \qquad (3)$$

where $Q_{us}$ is the number of stored charges and $Q_{uo}$ is the total number of photogenerated carriers [52]. This parameter provides a quantitative characterization of the long-term charge storage via steady photoconductance. For this parameter also, the enrichment of the hybrid nanocomposites with 5CB leads to improved results, namely, the charge storage efficiency increased obviously and the threshold value for intensity was decreased from 35 mW/cm$^2$ to 15 mW/cm$^2$. In Fig. 3(f), the dependencies of sensitivity on intensity for MoS$_2$/rGO and MoS$_2$/rGO-5CB samples are illustrated. Here, sensitivity (S) is calculated from the following expression:

$$S = i_{ph}/i_{dark}, \qquad (4)$$

where $i_{ph}$ is the photocurrent (difference between illumination current and dark current) [57]. The results indicate that the sensitivity of the MoS$_2$/rGO-5CB sample is nine times greater than that of MoS$_2$/rGO for the intensities in the range of 15–65 mW/cm$^2$. A possible explanation for the observed effect could be that

LCs induce channeling, which enhances both carrier mobility and electrical conductivity. It may facilitate the electrons' transfer from MoS$_2$ to rGO, which is expected to accelerate the separation of photoinduced electrons and holes.

Figs. 3(g and h) represent responsivity and specific detectivity parameters for the structure both with and without LC. Responsivity and specific detectivity are calculated according to the following expressions:

$$R = i_{ph}/I \cdot A, \tag{5}$$

$$D^* = R \cdot A^{\frac{1}{2}}/(2 \cdot e \cdot i_{dark})^{\frac{1}{2}}, \tag{6}$$

where $i_{ph}$ is the photocurrent, $I$ is the incident light intensity, and $A$ is the illuminated area on the sample, $i_{dark}$ is the dark current, and $e$ is the electron charge [54]. Here, the high responsivity for the MoS2/rGO-5CB could be due to the presence of trap states; however, further experimental and theoretical analysis would be needed to confirm this hypothesis and elucidate the exact role of the trap centers or to reveal other reasons. Moreover, the photoresponse of the samples displays a slight deterioration after 100 times of usage, indicating the reliability and robustness of the samples.

**Conclusion**

In summary, hydrothermally synthesized MoS$_2$/rGO hybrid nanocomposites were obtained in this work. Their opto-structural analysis was conducted using a wide range of methods including UV-Vis, PL, XPS, FTIR-ATR, Raman spectroscopy, XRD, TEM, SEM, and EDX. After synthesizing and characterizing the hybrid nanocomposites, 5CB with 5 % concentration was incorporated into the MoS$_2$/rGO for further tuning of its electro-optical properties. A detailed analysis of typical photodetector parameters, such as charge storage efficiency, sensitivity, responsivity, detectivity, and photocurrent decay time, was performed for both MoS$_2$/rGO and MoS$_2$/rGO-5CB hybrid nanocomposites. The findings demonstrate that the sensitivity of MoS$_2$/rGO-5CB is almost an order of magnitude higher than that of MoS$_2$/rGO. Besides, the photocurrent switching effect was observed when 5CB was incorporated into the hybrid nanocomposites. Moreover, the obtained photodetector characteristics showed an improvement of the hybrid nanocomposites with the addition of 5CB; specifically, the charge storage efficiency noticeably increased, and the threshold intensity value decreased from 35 mW/cm$^2$ to 15 mW/cm$^2$. Additionally, a high responsivity was observed for the MoS$_2$/rGO-5CB hybrid nanocomposites within the intensity range of 15 mW/cm$^2$ to 65 mW/cm$^2$. The achieved results offer novel routes for the interaction of nanocomposite 2D materials with LCs and their potential application in photodetectors and in photovoltaic devices and other optoelectronic devices or basic circuits.

**Supplementary Materials**

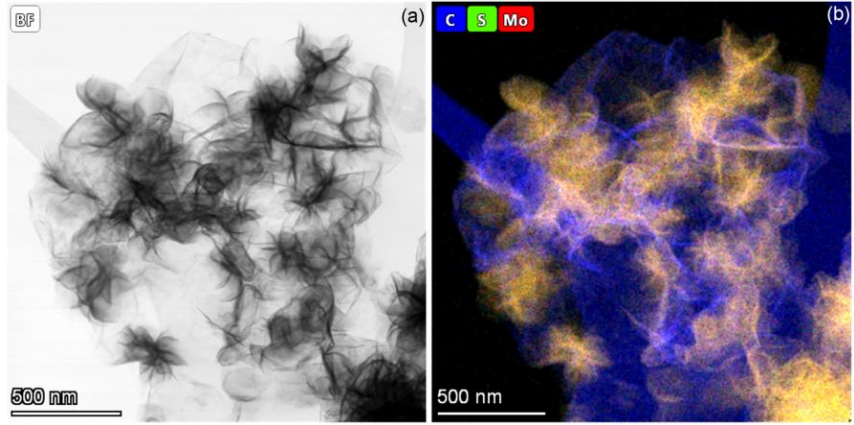

Fig. S1. TEM image of MoS$_2$/rGO morphology (a) an overlap of STEM-EDX chemical composition maps of C, S, and Mo net intensities.

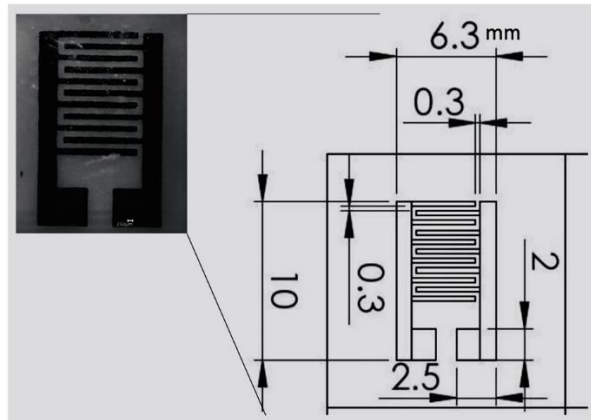

Fig. S2. Schematic view and sizes of interdigital electrodes used in the measurements.